\documentclass{article}
\usepackage{spconf,amsmath,graphicx}
\usepackage[backref=page]{hyperref}
\hypersetup{
	colorlinks=true,
	linkcolor=[rgb]{0.1796862745, 0.3134117647, 0.9127843137},
	citecolor=green,
	filecolor=magenta,      
	urlcolor=gray,
}
\usepackage{url}
\usepackage{amssymb,amsfonts}
\usepackage{algorithmic}
\usepackage{textcomp}
\usepackage{xcolor}
\usepackage{caption,subcaption}
\usepackage{tikz}
\usepackage{multirow}
\usepackage{diagbox}
\usepackage{hhline}
\usepackage{pgfplots}
\pgfplotsset{compat=1.17}
\usepackage{mathrsfs}
\usetikzlibrary{calc}
\usetikzlibrary{arrows}
\usetikzlibrary{positioning}
\usetikzlibrary{3d}

\graphicspath{{figures/}}

\title{PAON: A New Neuron Model using Padé Approximants}

\name{Onur Keleş$^{1,2}$, A. Murat Tekalp$^{1}$ \thanks{This work was supported by TUBITAK project 120C156. A. M. Tekalp also acknowledges support from Turkish Academy of Sciences (TUBA).}}
\address{$^{1}$Department of Electrical and Electronics Engineering, Koç University, İstanbul, Türkiye\\
	$^{2}$Codeway AI Research\\ 
	\texttt{\{okeles19, mtekalp\}@ku.edu.tr}}
	
\usepackage{fancyhdr}
\fancypagestyle{firstpage}{
	\fancyhf{} 
	\fancyhead[L]{Submitted to IEEE ICIP 2024} 
}

\begin{document}
	%
	\maketitle
	\thispagestyle{firstpage}
	\newcommand{\network}[1][s]{PadéNet#1}
	\newcommand{\layer}[1][s]{$\operatorname{\textit{PaLa}}$#1}
	\newcommand{\neuron}[1][s]{$\operatorname{\textit{Paon}}$#1}
	\newcommand{\mainact}{$\operatorname{GELU}$}
	\newcommand{\relu}{$\operatorname{ReLU}$}
	\newcommand{\pau}{$\operatorname{PAU}$}
	\newcommand{\first}[1]{\textcolor{red}{#1}}
	\newcommand{\second}[1]{\textcolor{blue}{#1}}
	\begin{abstract}
		Convolutional neural networks (CNN) are built upon the classical McCulloch-Pitts neuron model, which is essentially a linear model, where the nonlinearity is provided by a separate activation function. Several researchers have proposed enhanced neuron models, including quadratic neurons, generalized operational neurons, generative neurons, and super neurons, with stronger nonlinearity than that provided by the pointwise activation function. There has also been a proposal to use Padé approximation as a generalized activation function. In this paper, we introduce a brand new neuron model called Padé neurons (\neuron[s]), inspired by the Padé approximants, which is the best mathematical approximation of a transcendental function as a ratio of polynomials with different orders. We show that \neuron[s] are a super set of all other proposed neuron models. Hence, the basic neuron in any known CNN model can be replaced by \neuron[s]. In this paper, we extend the well-known ResNet to \network[] (built by \neuron[s]) to demonstrate the concept. Our experiments on the single-image super-resolution task show that \network[s] can obtain better results than competing architectures. 
	\end{abstract}
	
	\section{Introduction}
	\label{sec:intro}
	
	Convolutional neural networks (CNN) have become an accurate and reliable tool for solving many scientific and industrial problems. Although the rise of deep CNN is relatively recent \cite{krizhevsky2012imagenet}, the ideas behind their basic building block, the neuron model, are not new \cite{mcculloch1943logical, rosenblatt1957perceptron}. 
	
	The classical McCulloch-Pitts neuron, linearly combines each input element with different weights and then passes the result from a non-linear binary activation function. Later, many studies investigated more powerful activation functions to increase the capability of neurons leaving the linear part of the model intact. Although the rectified linear unit (\relu) \cite{jarrett2009best} remains as the most popular choice following \cite{krizhevsky2012imagenet}, other variants, such as leaky \relu\ \cite{maas2013rectifier}, Gaussian error linear unit \cite{hendrycks2016gaussian} and sigmoid linear unit \cite{elfwing2018sigmoid} are also commonly used. Noting that these are pre-determined and hand-crafted non-linearities, Molina et al. \cite{molina2019pade} proposed to learn the activation function for each layer via Padé approximation initializing the coefficients from a decided non-linearity.
	
	Based on the idea that a neuron model should not be limited to a pointwise nonlinearity introduced by the activation, new inherently nonlinear neuron models have been proposed. In this thread of research, quadratic neurons \cite{cheung1991rotational, milenkovic1996annealing, xu2022quadralib, chen2023expressivity} propose to operate on both first and second powers of their inputs. Generalized operational perceptrons \cite{kiranyaz2020operational} propose to replace the weighted linear combination and addition operations in the classical neuron model with different mathematical functions. Generative neurons \cite{kiranyaz2021self} are inspired by the Taylor series expansion for polynomial approximation of arbitrary nonlinear functions by operating on the higher order powers of the input, and were applied to different image processing tasks \cite{kelecs2021self, yilmaz2021self}. Super neurons \cite{kiranyaz2023super} aim to expand the receptive field of generative neurons via learnable shifts applied to convolution kernels. More details are in Section \ref{sec:rel_work}.
	
	Inspired by these works, this paper presents a new and more powerful inherently nonlinear neuron model called \neuron[] using Padé approximation of nonlinear functions. \neuron[s] can learn any nonlinear function as a ratio of polynomials, which is a more powerful alternative to Taylor series expansion. We show that \neuron[s] are a super set of other known neuron models and can replace the classic neuron model in any CNN. 
	\begin{figure*}[th]
		\centering
		\includegraphics[width=0.83\textwidth]{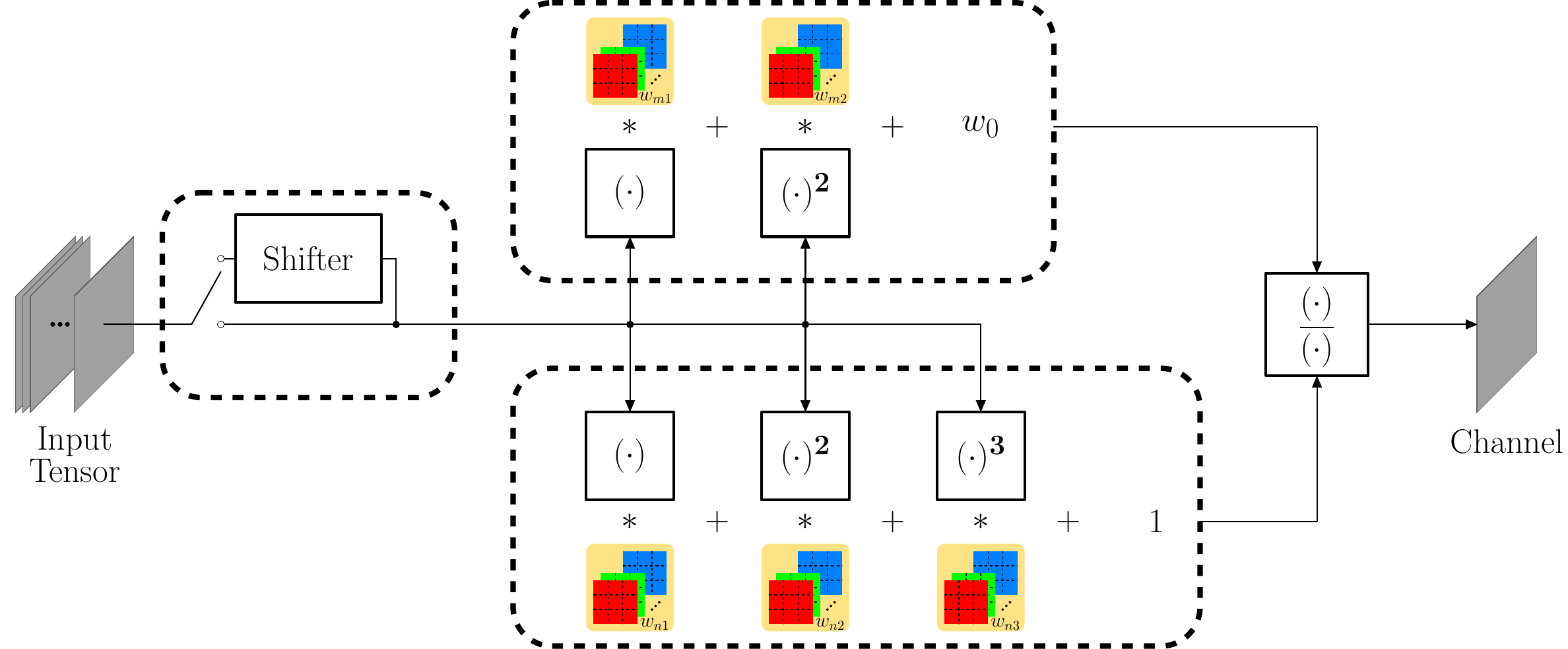}
		\caption{Illustration of a Padé neuron (\neuron[]) for $[M/N] = [2/3]$, where $w_{0}$ is bias for numerator, $ (\cdot)^{k} $ takes $k^{\text{th}}$ power of the~input in element-wise manner, $\frac{(\cdot)}{(\cdot)}$ implements either Eq.~\eqref{eq:pade_neuron_abs} or \eqref{eq:pade_neuron_smooth}, and $ \mathbf{\ast} $ is convolution. The shifter module shifts the~input features.}
		\vspace{-10pt}
		\label{fig:pade_neuron}
	\end{figure*}
	
	\section{Related Works}
	\label{sec:rel_work}
	
	There have been several attempts to define more powerful generalized activations or inherently non-linear neuron models, which are discussed in detail below: 
	
	{\bf Quadratic neurons.} Quadratic neurons define a nonlinear relationship between its inputs and outputs by operating on the input $x$ as well as the square of the input $x^2$ given by
	\begin{equation}
		\label{eq:general_quadratic_neuron}
		f(x) = A(x^{2}) + B(x),
	\end{equation}
	where $A$ is a quadratic function of $x$, $B$ is linear in $x$. Here, the bias is omitted for simplicity. 
	
	This general formulation was employed in various studies. Cheung and Leung \cite{cheung1991rotational} used $x^{\text{T}}w_{1}x + w_{2}x$, authors of \cite{milenkovic1996annealing} modified the second term as $w_{2}x^{2}$. The study \cite{bu2021quadratic} obtained quadratic expression via multiplying two filtered inputs as $(w_{1}x)\odot(w_{2}x)$, where $\odot$ is element-wise (Hadamard) product. \cite{xu2022quadralib} added $w_{3}x$ to the previous expression, and \cite{chen2023expressivity} used low-rank approximation to calculate quadratic terms. Unlike the quadratic neurons, the proposed \neuron[s] do not restrict themselves to only second order polynomials.
	
	{\bf Generalized Operational Perceptrons.} Quadratic neurons are only able to express functions that are second order at most. Kıranyaz et al. \cite{kiranyaz2020operational} introduce a new neuron model, which replace linear scaling of input with weights and addition of the results with a selected set of complex mathematical operations. Generalized operational perceptrons can apply various functions as their \textquotedblleft nodal\textquotedblright\ operator, such as exponentiation, taking sinusoidal and so on, as well as scaling by linear weights as in a common neuron. Moreover, its \textquotedblleft pool\textquotedblright\ operation (addition in a regular neuron) can be some other appropriate operation such as median operator. However, it is computationally very expensive both to choose those operations and apply them since they take more resources compared to addition and multiplication. Moreover, the choices are very architecture-dependent; if, say, a structural change is desired to be made by adding one more layer, another extensive search has to be done again to find the specific operations. 
	
	{\bf Generative Neurons.} Noticing that huge computation necessity, the study \cite{kiranyaz2021self} proposes generative neurons. They basically try to approximate the required mapping function by truncated Taylor series expansion around the point $0$; i.e., they apply Maclaurin series expansion up to pre-determined order. By this way, generative neurons aim to work around the computation burden of the generalized operational perceptrons while still trying to be able to catch an equivalent non-linearity. However, since they are linear combination of different positive orders of the input, which can go out of the range for safe computation zone, and Taylor series approximation is the best around a specific point and worse on other further points, the output of generative neurons \cite{kiranyaz2021self} had to be limited by $\tanh$ activation, which is known to be a source of vanishing gradients, thus, impediments the training of deep models. On the contrary, in \neuron[s], higher ordered approximations can be calculated as a ratio of two polynomials. Thanks to this property, in many cases \network[s] do not require any limiting activation and can benefit from the common non-linearities that are known to overcome the vanishing gradient problem. Moreover, for a given approximation order, Padé approximant can follow the target transcendental function closer than the Taylor series expansion around a point \cite{baker1996pade}. Thus, for the same amount of non-linearity, \neuron[s] serve as a more efficient way.
	
	{\bf Super Neurons.} The generative neuron has a local receptive field; i.e., all kernels for different powers pull information from the same location on a feature map. Superneurons \cite{kiranyaz2023super} introduce shifts, which are randomly initialized and optimized via back propagation during training. In contrast, \neuron[s] learn the shifts from the data via the $\operatorname{Shifter}$ module.
	
	{\bf Padé Activation Unit (PAU).} The study \cite{molina2019pade} proposes to use Padé approximant as an activation, so called Padé activation unit (\pau). They pre-determine the orders for rational polynomials as well as some starting coefficients for preferred activation as an initial non-linearity. In \pau, however, the activation function is learned for a whole layer and tends to have the same shape as the non-linearity whose Padé approximation is used as a starting point for coefficients. In \network[s], every single element in each neuron learns its own approximation. Thus, a single neuron in a layer with $k\times k$ kernel actually learns $k^2$ different Padé approximant as a ratio of two polynomials. Thus it brings higher degrees of freedom to our choice, and provides element-wise non-linearity to each kernel addition to the one coming from the activation function.
	
	\begin{figure*}[ht]
		\centering
		\begin{minipage}{0.68\textwidth}
			\centering
			\includegraphics[width=\textwidth]{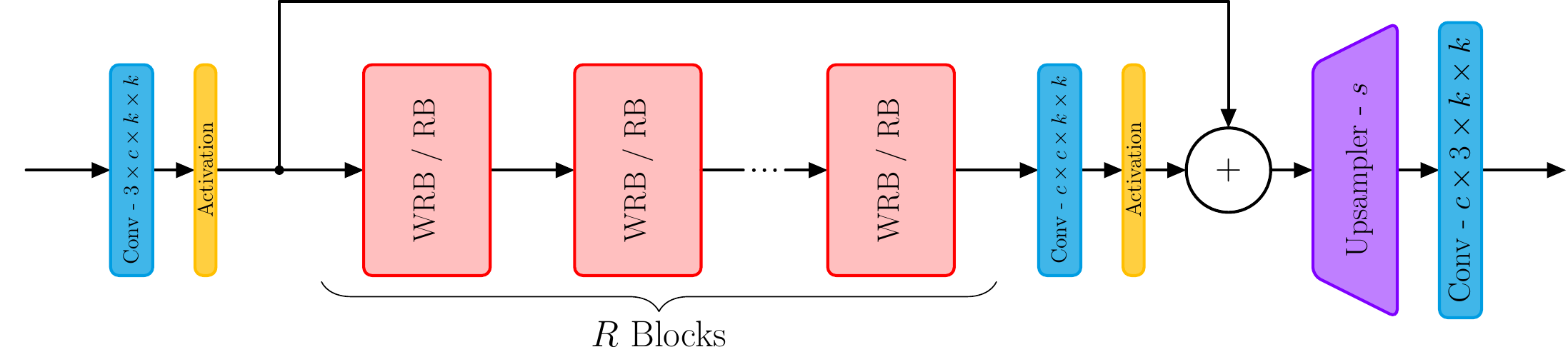}
			\subcaption{}
			\label{fig:sr_arch}
		\end{minipage}
		\begin{minipage}{0.3\textwidth}
			\centering
			\vspace{5.pt}
			\includegraphics[width=\textwidth]{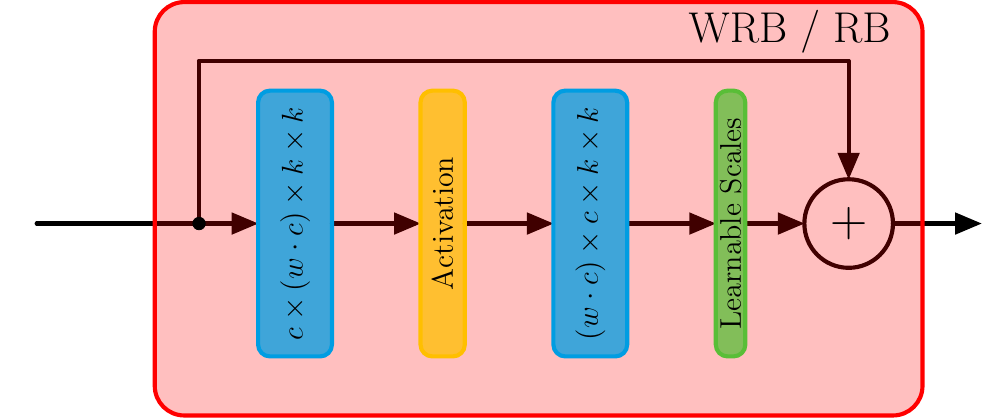}
			\vspace{-1pt}
			\subcaption{}
			\label{fig:resblock}
		\end{minipage}
		\caption{(\subref{fig:sr_arch}): The architecture for the super-resolution experiments. (\subref{fig:resblock}): Normal and wide residual block structure. For wide residual block (WRB), $w$ is bigger than $1$ while for normal block (RB), it is $1$.}
	\end{figure*}
	
	\section{\textit{PAON}: Padé Approximant Neuron Model}
	\label{sec:pade_networks}
	
	Padé approximant is the best approximation of a transcendental function by a ratio of two polynomials with given orders. Let $f_{[M/N]}(x)$ denote an approximation expression for function $f$ with $M$- and $N$-degree polynomials in numerator and denominator, respectively. Then, the Padé approximant for the function $f$ can be written as
	\begin{equation}
		\label{eq:poly_ratio}
		f_{[M/N]}(x) = \dfrac{P_{M}(x)}{Q_{N}(x)} = \left.{\sum_{k=0}^{M}a_kx^k}\middle/{\sum_{k=0}^{N}b_kx^k}\right.
	\end{equation}
	where $a_k$s and $b_k$s are the coefficients of polynomials in numerator and denominator, respectively. Conventionally, Padé approximant coefficients are normalized such that $b_0 = 1$. Thus, if we rewrite the expression,
	\begin{equation}
		\label{eq:pade_approx}
		f_{[M/N]}(x) = \dfrac{\displaystyle\sum_{k=0}^{M}a_kx^k}{1 + \displaystyle\sum_{k=1}^{N}b_kx^k} =  \dfrac{a_{0} + \displaystyle\sum_{k=1}^{M}a_kx^k}{1 + \displaystyle\sum_{k=1}^{N}b_kx^k}. %
	\end{equation}
	If we think $a_k$s and $b_k$s as convolutional kernels and $a_0$ as the~bias term, then we have a neuron model: 
	\begin{equation}
		\label{eq:vanilla_pade_neuron}
		f_{[M/N]}(x) = \dfrac{P_{M}(x)}{Q_{N}(x)} = \dfrac{w_{0} + \displaystyle\sum_{k=1}^{M}w_{mk}\ast x^k}{1 + \displaystyle\sum_{l=1}^{N}w_{nl}\ast x^l},
	\end{equation}
	where $w_{mk}$ and $w_{nl}$ are kernels of numerator and denominator for the input of order $k$ and $l$, respectively, and $w_0$ is the bias term for numerator. Fig. \ref{fig:pade_neuron} shows the workflow of a Padé neuron (\neuron[]) for $[M/N] = [2/3]$.
	
	{\bf Singularity of Pade Approximants.} One thing to note about this neuron model is that the denominator has the potential to be equal or very close to $0$. Although the weights can be initialized to prevent this at the beginning, learning of coefficients with gradient descent does not guarantee it to remain away from zero. To mathematically ensure that the divisor is always nonzero, we propose two variants of the Padé neurons. First, we take the absolute value of each individual power in the denominator so that it is guaranteed to divide each element in the numerator by a number greater than or equal to $1$. So, the final expression for the Padé neuron with absolute value, \neuron[]-$A$, becomes 
	\begin{equation}
		\label{eq:pade_neuron_abs}
		f_{[M/N]}(x) = \dfrac{w_{0} + \displaystyle\sum_{k=1}^{M}w_{mk}\ast x^k}{1 + \displaystyle\sum_{l=1}^{N}\left|w_{nl}\ast x^l\right|}.
	\end{equation}
	The second variant is inspired by the work \cite{beckermann1997diagonal}. Although this method was proposed for diagonal Padé approximants, i.e., $M=N$, we  observe that it can be used for $\lvert M-N\rvert=\{0,1\}$ without any further modification. This smoothed variant of the Padé neuron, \neuron[]-$S$, can be written as
	\begin{equation}
		\label{eq:pade_neuron_smooth}
		f_{[M/N]}(x) = \dfrac{Q_{N}(x)P_{M}(x) + Q_{N-1}(x)P_{M-1}(x)}{Q_{N}^{2}(x) + Q_{N-1}^{2}(x)},
	\end{equation}
	where $P_{M}(x)$ and $Q_{N}(x)$ are as defined in Eq.~\eqref{eq:vanilla_pade_neuron}.
	
	Eqs. \eqref{eq:pade_neuron_abs} and \eqref{eq:pade_neuron_smooth} tell that every kernel element in a \neuron[] adapts itself so that each weight group in a kernel learn their specific Padé approximant. It brings more non-linearity as it introduces the higher orders of features to the model both as the numerator and denominator. Moreover, thanks to its being a ratio of polynomials, it behaves more stable for higher order approximations, thus, it does not need to limit its output with bounded activation functions such as $\tanh$. 
	
	{\bf Shifter.} $\operatorname{Shifter}$ module consists of averaging, $1\times1$ convolution and a non-linear activation function together with some viewing operations for shape consistency. When it takes a negative number as shift parameter, it is deactivated. When the shift parameter $b$ is a positive integer, it performs gradient-based optimization to find the best shift in the range $[-b, b]$, and when $b=0$, it computes the best shift for each channel without any restriction. The convolution weights and bias in $\operatorname{Shifter}$ module are initialized as zero to make sure that the module learns the amount of shift only when it does not hurt the performance.
	
	{\bf Paons as Superset of Existing Neuron Models.} Padé neurons are the super set of the aforementioned neuron models. For $M=1$, $N=0$ and $\operatorname{Shifter}$ is not active, the \neuron[] becomes an ordinary neuron\footnote{In Padé approximant, when $N$ is $0$, the denominator becomes $1$.}. When $M=2$, $N=0$, it shows the properties of a quadratic neuron. For $M\geq2$ and $N=0$, they behave as a generative neuron, and when the $\operatorname{Shifter}$ branch is activated, it behaves as a super neuron with improved performance since it learns the effective shifts from the data. Thus, \neuron[s] can easily replace any neuron model in a convolutional network.
	
	\section{Experiments}
	\label{sec:experiments}
	
	\subsection{Architecture}
	\label{subsec:architecture}
	
	We show the performance of \network[s] on the single-image super resolution problem. The basic architecture chosen for this task is a widely used one since the seminal paper \cite{ledig2017photo}. In this architecture, a single feature extraction layer is followed by a series of blocks for residual feature refinement. For simplicity, we chose residual \cite{he2016deep} and wide residual blocks \cite{zagoruyko2016wide} with scaled residuals \cite{szegedy2017inception} as our feature refinement blocks. The initial features are added back to the refined residual features. The sum is processed by a feature upsampler module, which contains a layer, an activation, and a PixelShuffler layer \cite{shi2016real}. The employed architecture is shown in Fig. \ref{fig:sr_arch}, and Fig. \ref{fig:resblock} shows the structure of a residual and wide residual block. The learnable scaler layer for each output channel is initialized from $0.1$.
	
	We check the performance of \network[s] by comparing various architectures: a wide residual network network composed of convolutional layers and \mainact\ activation (called ResNet), a network with convolutional layer and \pau\ activation (called \pau-Net), a SelfONN, a SuperONN, and \network[]. In all of the models, we keep the convolutions at the initial feature extractor, ath the end of feature refinement, and the final image constructor part (upsampler and final layer) the same degrees as $[1/0]$ to be able to keep track of the number of parameters.  
	
	\begin{table}
		\centering
		\caption{PNSR scores on DIV2K$\times2$ validation set. \textquotedblleft FL\textquotedblright, \textquotedblleft LL\textquotedblright\ and \textquotedblleft AL\textquotedblright\ denote first layer, last layer and all layers are \layer[], respectively. All values are calculated using \neuron[]-$S$, except for the \neuron[]-$A$ column.}
		\resizebox{0.48\textwidth}{!}{
			\begin{tabular}{||c|c||c|c||c|c|c||}
				\hline
				No Shift & Shift & \neuron[]-$A$ & \neuron[]-$S$ & FL & LL & AL \\\hline
				$34.32$  & $\mathbf{34.36}$ & $34.35$ & $\mathbf{34.36}$ & $34.29$ & $34.30$ & $\mathbf{34.36}$ \\\hline
			\end{tabular}
		}
		\label{tbl:arch_choice}
		\vspace{-5pt}
	\end{table}
	
	Before the final comparison, we investigate the performance of \network[] in different setups. We check which \neuron[] is better, whether or not the $\operatorname{Shifter}$ is helpful, and which layers in RB to be converted into Pade neuron layers (\layer[]). The results are shown in Table \ref{tbl:arch_choice}. According to those metrics, we continue with \neuron[]-$S$ with $\operatorname{Shifter}$ is activated in both of the layers in the residual block. 
	
	Image boundaries are processed in 3x3 convolutions and $\operatorname{Shifter}$ module according to circular extension \cite{tekalp1985boundary}.
	
	For the final comparison, architecture details are given in Table \ref{tbl:archs_overview}. Note that SelfONN and SuperONN have $\tanh$ activations. In the experiments with \mainact, those two networks became unstable, so we used $\tanh$ in residual blocks as proposed in their studies.
	
	\begin{table}[h]
		\centering
		\caption{Model configurations. Degrees denote the degree of numerator/denominator polynomials $[M/N]$. The degree for \pau-Net is the degree of the \pau\ activation. RB and WRB denote residual block and wide RB, respectively. }
		\resizebox{0.48\textwidth}{!}{
			\begin{tabular}{|c||c|c|c|c|}
				\hline
				\multicolumn{1}{|c||}{} & ResNet & \pau-Net & (Self/Super)ONN & \network[] \\\hhline{|=||=|=|=|=|}
				Degrees & $[1/0]$ & $[7/6]$ & $[3/0]$ & $[2/1]$ \\\hline
				Activation & \mainact & \pau & $\tanh$ & \mainact \\\hline
				Block Type & WRB & WRB & RB & RB \\\hline
				Blocks, $R$ & $3$ & $3$ & $3$ & $3$ \\\hline
				Channels & $48$ & $48$ & $48$ & $48$ \\\hline
			\end{tabular}
		}
		\label{tbl:archs_overview}
		\vspace{-10pt}
	\end{table}
	
	\subsection{Training Details}
	\label{subsec:train}
	
	For the training, we use DF2K dataset \cite{lim2017enhanced} as it has more images compared to DIV2K \cite{Agustsson_2017_CVPR_Workshops, Timofte_2017_CVPR_Workshops}. The models are trained on $64\times64$ patches scaled to $[-1, 1]$ range with $25$ batch size for $5\times10^{5}$ iterations to perform on $\times2$ and $\times4$ super-resolution. The data are augmented with random rotation, horizontal and vertical flip, and color channel shuffling. Also, in the experiments, we noted that adding a small amount of Gaussian noise during training improves the validation score of the network. Therefore, we add Gaussian noise with $40\text{ dB}$ SNR into the cropped patches. The model tries to minimize the loss function with $\alpha=1.5$ and $c=2$, proposed in \cite{barron2019general}. We use Adan optimizer \cite{xie2022adan} with $10^{-3}$ learning rate and cosine annealing scheduler \cite{loshchilov2016sgdr} until the learning rate becomes $10^{-6}$. The best model is saved with respect to its validation PSNR on the DIV2K validation set.
	
	For comparison, the standard sets in super-resolution that are BSD100 \cite{martin2001database}, Manga109 \cite{matsui2017sketch}, Set5 \cite{bevilacqua2012low}, Set14 \cite{zeyde2012single} Urban100 \cite{huang2015single} are used. For all of the compared models, PSNR, SSIM and LPIPS metrics are reported to compare the performance. PSNR is calculated from RGB images \cite{keles2021computation}. For SSIM \cite{wang2004image}, Y channel of YCbCR image is used. LPIPS \cite{zhang2018unreasonable} results are reported from both AlexNet \cite{krizhevsky2012imagenet} and VGG \cite{simonyan2014very}.
	
	\subsection{Results and Discussion}
	\label{subsec:results}
	
	\begin{table*}[h]
		\centering
		\caption{Quantitative comparison. The top two scores in each cell are PSNR($\uparrow$) and SSIM($\uparrow$), and the bottom two are LPIPS($\downarrow$) calculated via AlexNet and VGGNet, respectively. The best and second best scores for each dataset are shown in \textcolor{red}{red} and \textcolor{blue}{blue}.} \vspace{-5pt}
		\resizebox{\textwidth}{!}{
			\begin{tabular}{|c||c|c|c|c|c||c|c|c|c|c|} 
				\cline{2-11}
				\multicolumn{1}{c|}{} & \multicolumn{5}{c||}{$\times2$} & \multicolumn{5}{c|}{$\times4$} \\\hline
				{} & ResNet & \pau-Net & SelfONN & SuperONN & \network[] & ResNet & \pau-Net & SelfONN & SuperONN & \network[] \\\hhline{|=||=|=|=|=|=||=|=|=|=|=|}
				\multirow{2}{*}{BSD100} 
				& $30.64/0.8903$ & $30.63/0.8903$ & $\second{30.65}/0.8904$ & $\second{30.65}/0.8905$ & $\first{30.68}/\first{0.8909}$ 
				& $\second{26.13}/\second{0.7133}$ & $26.11/0.7127$ & $26.12/0.7125$ & $\second{26.13}/0.7131$ & $\first{26.16}/\first{0.7138}$ \\
				
				& $0.1513/0.1580$ & $0.1516/\second{0.1576}$ & $\second{0.1508}/0.1584$ & $\first{0.1501}/0.1578$ & $0.1516/\first{0.1574}$
				& $\first{0.3877}/0.3434$ & $0.3892/0.3434$ & $0.3892/\second{0.3432}$ & $0.3893/0.3444$ & $\second{0.3884}/0.3439$ \\\hline
				\multirow{2}{*}{Manga109} 
				& $35.60/\second{0.9724}$ & $35.60/0.9723$ & $35.62/\second{0.9724}$ & $\second{35.63}/\second{0.9724}$ & $\first{35.70}/\first{0.9727}$ 
				& $\second{28.16}/\second{0.8924}$ & $28.11/0.8912$ & $28.14/0.8908$ & $\second{28.16}/0.8922$ & $\first{28.28}/\first{0.8941}$ \\
				
				& $0.0242/0.0568$ & $0.0241/\first{0.0561}$ & $\second{0.0235}/\second{0.0566}$ & $0.0236/0.0567$ & $\first{0.0234}/0.0569$
				& $0.1166/0.1720$ & $0.1172/\second{0.1696}$ & $0.1168/\first{0.1685}$ & $\first{0.1156}/0.1728$ & $\second{0.1161}/0.1726$ \\\hline
				\multirow{2}{*}{Set5} 
				& $35.50/0.9548$ & $35.51/0.9548$ & $\first{35.57}/\second{0.9550}$ & $\second{35.54}/0.9549$ &  $\second{35.54}/\first{0.9551}$ 
				& $29.78/0.8775$ & $29.81/0.8771$ & $29.85/0.8774$ & $\second{29.88}/\second{0.8776}$ & $\first{29.90}/\first{0.8788}$ \\
				
				& $0.0591/0.0964$ & $0.0591/\first{0.0955}$ & $0.0585/\second{0.0957}$ & $\second{0.0582}/0.0964$ & $\first{0.0580}/0.0958$
				& $0.1808/0.2184$ & $\second{0.1807}/\second{0.2172}$ & $\first{0.1799}/\first{0.2156}$ & $0.1812/0.2185$ & $0.1812/0.2192$ \\\hline
				\multirow{2}{*}{Set14} 
				& $30.99/\second{0.9074}$ & $30.97/0.9073$ & $31.11/\second{0.9074}$ & $\second{31.12}/0.9073$ & $\first{31.14}/\first{0.9080}$ 
				& $26.29/0.7594$ & $26.23/0.7594$ & $26.32/0.7591$ & $\second{26.35}/\second{0.7598}$ & $\first{26.37}/\first{0.7604}$ \\
				
				& $0.0985/0.1462$ & $0.0994/\second{0.1457}$ & $0.0985/0.1466$ & $\second{0.0984}/0.1462$ & $\first{0.0981}/\first{0.1456}$
				& $\first{0.2947}/0.3090$ & $0.2952/\second{0.3080}$ & $\second{0.2951}/0.3085$ & $0.2956/0.3092$ & $0.2958/\first{0.3078}$ \\\hline
				\multirow{2}{*}{Urban100} 
				& $29.76/0.9148$ & $29.75/0.9147$ & $\second{29.82}/\second{0.9153}$ & $29.80/\second{0.9153}$ & $\first{29.90}/\first{0.9165}$ 
				& $24.27/0.7598$ & $24.23/0.7582$ & $24.26/0.7585$ & $\second{24.30}/\second{0.7608}$ & $\first{24.35}/\first{0.7624}$ \\
				
				& $0.0763/0.1198$ & $0.0767/0.1198$ & $0.0757/0.1193$ & $\second{0.0750}/\second{0.1191}$ & $\first{0.0742}/\first{0.1183}$
				& $0.2540/0.2944$ & $0.2559/0.2946$ & $0.2559/\first{0.2940}$ & $\second{0.2517}/\second{0.2942}$ & $\first{0.2515}/\first{0.2940}$ \\\hline
			\end{tabular}
		}
		\vspace{-10pt}
		\label{tbl:quant_results}
	\end{table*}
	
	The quantitative resulst are shown in Table~\ref{tbl:quant_results}. It can be seen that \network[] surpasses all the models in every fidelity metrics, PSNR and SSIM, for nearly all the datasets. The PSNR difference with wide ResNet can be as much as $0.15\text{ dB}$. This indicates that although the number of parameters are close to each other, a model equipped with function approximation capability can serve better in the signal reconstruction. Moreover, comparison with \pau-Net shows that increasing the approximation capability of the network via introducing the Padé approximants to the each kernel element increases the expressive capability of the models. Finally, using Padé approximant rather than Taylor series expansion in the neurons proves to be a better strategy, thanks to the more accurate approximation capability of former over the latter. This shows that the networks built with \neuron[] can harness the most of the input information and express the desired function that a neural network tries to approximate in a better way.
	
	\begin{figure}[t]
		\centering
		\begin{tikzpicture}
			\node[anchor=south west,inner sep=1] (img) at (0,0) {\includegraphics[width=0.48\textwidth]{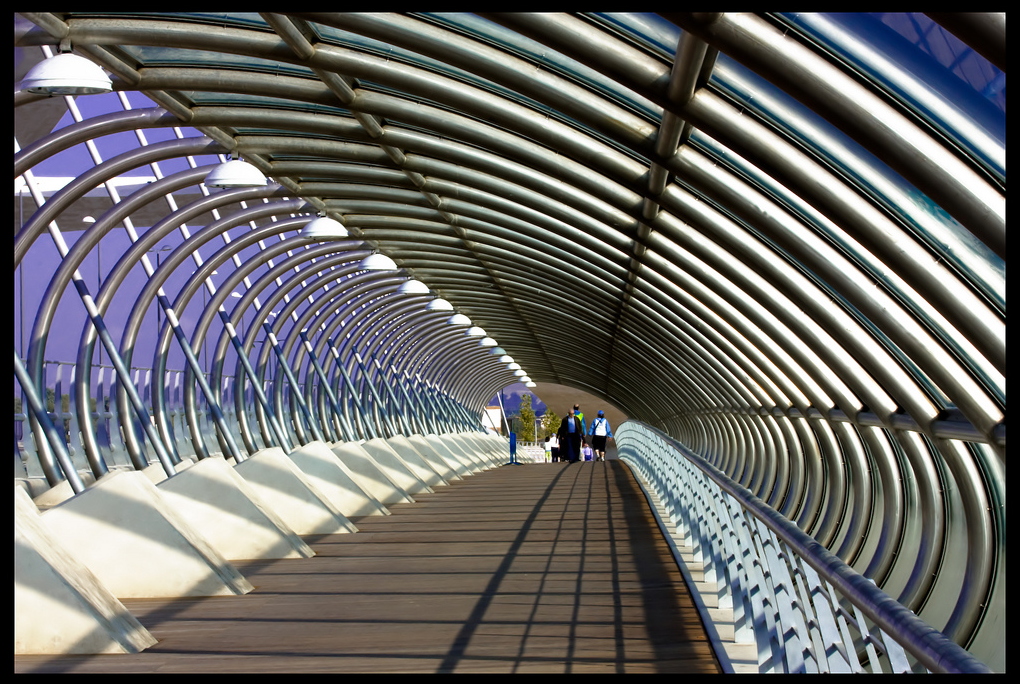}};
			\begin{scope}[x={(img.south east)},y={(img.north west)}]
				\draw[red,thick] (0.4559,0.4737) rectangle (0.5049,0.5102); 
			\end{scope}
		\end{tikzpicture}
		\begin{subfigure}{0.19\columnwidth}
			\includegraphics[width=\columnwidth]{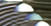}
			\label{fig:x2_resnet_crop}
		\end{subfigure}
		\begin{subfigure}{0.19\columnwidth}
			\includegraphics[width=\columnwidth]{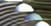}
			\label{fig:x2_pau_crop}
		\end{subfigure}
		\begin{subfigure}{0.19\columnwidth}
			\includegraphics[width=\columnwidth]{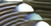}
			\label{fig:x2_self_crop}
		\end{subfigure}
		\begin{subfigure}{0.19\columnwidth}
			\includegraphics[width=\columnwidth]{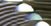}
			\label{fig:x2_superonn_crop}
		\end{subfigure}
		\begin{subfigure}{0.19\columnwidth}
			\includegraphics[width=\columnwidth]{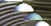}
			\label{fig:x2_pade_crop}
		\end{subfigure}
		\vspace{-10pt}
		\caption{Visual comparison for $\times2$ SR on \texttt{img\_058.png} from Urban100 dataset. Crops from left to right are outputs of ResNet, \pau-Net, SelfONN, SuperONN and \network[].}
		\label{fig:x2_visual}
	\end{figure}
	
\begin{figure}[ht]
		\centering
		\begin{tikzpicture}
			\node[anchor=south west,inner sep=1] (img) at (0,0) {\includegraphics[width=0.48\textwidth]{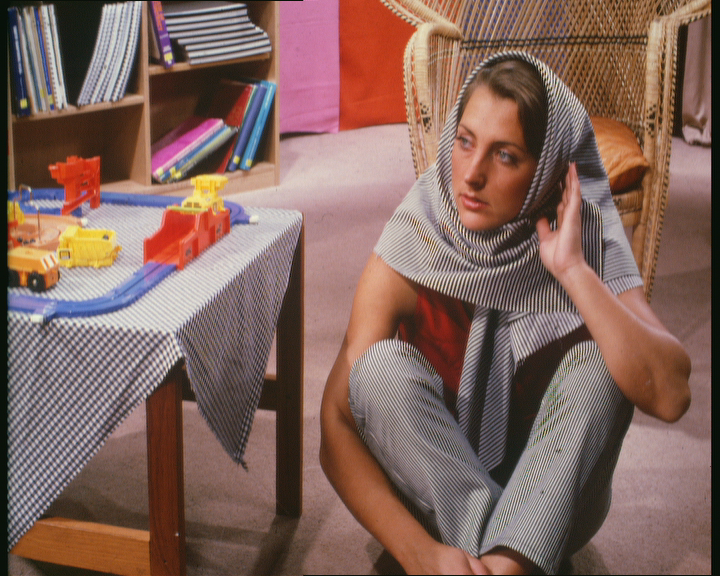}};
			\begin{scope}[x={(img.south east)},y={(img.north west)}]
				\draw[red,thick] (0.0208,0.0972) rectangle (0.1667,0.3403); 
			\end{scope}
		\end{tikzpicture}
		\begin{subfigure}{0.19\columnwidth}
			\includegraphics[width=\columnwidth]{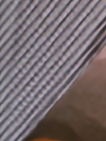}
			\label{fig:x4_resnet_crop}
		\end{subfigure}
		\begin{subfigure}{0.19\columnwidth}
			\includegraphics[width=\columnwidth]{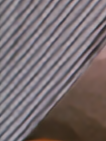}
			\label{fig:x4_pau_crop}
		\end{subfigure}
		\begin{subfigure}{0.19\columnwidth}
			\includegraphics[width=\columnwidth]{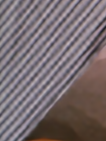}
			\label{fig:x4_self_crop}
		\end{subfigure}
		\begin{subfigure}{0.19\columnwidth}
			\includegraphics[width=\columnwidth]{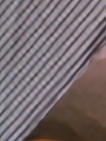}
			\label{fig:x4_superonn_crop}
		\end{subfigure}
		\begin{subfigure}{0.19\columnwidth}
			\includegraphics[width=\columnwidth]{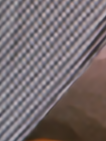}
			\label{fig:x4_pade_crop}
		\end{subfigure}
		\vspace{-10pt}
		\caption{Visual comparison for $\times4$ SR on \texttt{barbara.png} from Set14 dataset. Crops from left to right are outputs of ResNet, \pau-Net, SelfONN, SuperONN and \network[].}
		\label{fig:x4_visual}
		\vspace{-15pt}
	\end{figure}

	Figs. \ref{fig:x2_visual} and \ref{fig:x4_visual} show the qualitative results of the models. Since the $\times2$ SR is a relatively easier problem than $\times4$ SR, the nuances are harder to detect. In Fig. \ref{fig:x2_visual}, it can be seen that the cropped region has some high frequency content. Inspecting the images, it can be claimed that the most successful model to reconstruct those details is \network[]. The other models either cause aliasing, or fail to reconstruct a straight line (in case of SelfONN). The difference is clearer in Fig. \ref{fig:x4_visual}. The table cover normally has square patterns, which all models fail to reconstruct. But \network[] is again the closest one to bring the perpendicular lines into the image. Although the Resnet also shows them, the output of \network[] has more details.
	
	\section{Conclusion and Future Work}
	\label{sec:conclusion}
	
	In this paper, we introduce a new neuron model called Pade approximant neurons, or \neuron[] in short. It enhances the~non-linear capability of a regular convolutional neuron via utilizing the higher order polynomials and Padé approximants on each of the kernel element. Its construction makes it the super set of the previously proposed neuron models such as quadratic neurons, generative neurons and super neurons. We show an application of \neuron[s] on the singe-image super-resolution problem. Quantitative results show that (1) it outperforms the regular neuron model thanks to its highly non-linear nature and (2) \neuron[] surpasses the recently proposed generative and super neurons thanks to its better approximation capability and learnable shifter module.
	
	As future work, we intend to check its performance on different image-related tasks. Also, the increase on the $\operatorname{Shifter}$ performance would be the next step to further increase the neuron performance.
	\bibliographystyle{IEEEbib}
	\bibliography{refs}
	
\end{document}